# New clues on the extended HeII ionization in IZw18 from GTC/MEGARA and JWST/MIRI

A. Arroyo-Polonio,[1] C. Kehrig,[1,2] J.M. Vílchez,[1] J. Iglesias-Páramo,[1,3] E. Pérez-Montero,[1]
S. Duarte Puertas,[1,4] J. Gallego,[5] and D. Reverte[6]

[1]*Instituto de Astrofísica de Andalucía, CSIC, Apartado de correos 3004, 18080 Granada, Spain*
[2]*Observatório Nacional/MCTIC, R. Gen. José Cristino, 77, 20921-400, Rio de Janeiro, Brazil*
[3]*Centro Astronómico Hispano en Andalucía, Observatorio de Calar Alto, Sierra de los Filabres, 04550 Gérgal, Spain*
[4]*Departamento de Física Teórica y del Cosmos, Universidad de Granada, Granada, Spain*
[5]*Departamento de Física de la Tierra y Astrofísica, Universidad Complutense de Madrid*
[6]*Instituto de Astrofísica de Canarias, C/ Vía Láctea s/n, La Laguna, Tenerife, Spain*

## ABSTRACT

IZw18 is one of the lowest-metallicity star-forming galaxies known at z∼0, considered a unique local analogue of the first galaxies. The origin of its hard ionizing continuum, expected to be a common feature in the early Universe and traced by HeII emission lines, remains intensely debated and challenging to explain. Here we combine optical (GTC/MEGARA) and mid-infrared (JWST/MIRI) integral field spectroscopic observations for IZw18 to shed new light on the high-ionization phenomenon. This letter reports the first detection of the high-ionization [NeV]14.32 $\mu$m line in IZw18. Its emission is spatially extended and coincident with the HeII peak, revealing the presence of highly energetic ionizing sources that surpass mechanisms previously proposed on the basis of HeII alone. Our kinematic studies highlight that the HeII$\lambda$4686-emitting gas displays higher velocity dispersions and a different velocity pattern compared to the H$\beta$ emission, suggesting the presence of energetic processes such as shocks or stellar-driven feedback. Additionally, integrated spectra show asymmetric blueshifted profiles in the HeII$\lambda$4686 line, possibly indicating early-stage stellar-driven outflows potentially facilitating future ionizing photon leakage. Our spatial analysis also reveals differences in structure between the emission of H$\beta$ and HeII$\lambda$4686, with the HeII$\lambda$4686 peak offset by a projected distance of 140 pc from the peak H$\beta$ emission. This indicates distinct locations for the most extreme ionizing sources compared to moderate ionizing sources. Our findings underscore the complex interplay of physical processes in extremely metal-poor environments with high-ionized gas, offering new insights into the conditions prevailing in the early galaxies.

## 1. INTRODUCTION

The detection of nebular (narrow) HeII recombination lines, particularly at $\lambda$1640 Å (UV) and $\lambda$4686 Å (optical), in star-forming (SF) galaxies signals the presence of hard radiation (Energy ≥ 4 Ryd ∼ 54.4 eV). The nature and origin of such high-ionization lines, especially in metal-poor SF systems, remains unexplainable and actively debated in the literature (e.g., D. R. Garnett et al. 1991; C. Kehrig et al. 2015; L. M. Oskinova &

Email: aarroyo@iaa.es

D. Schaerer 2022). This persistent ambiguity about the sources capable of doubly ionizing helium hampers our interpretation of the nebular He II emission lines in high-redshift galaxies and, by extension, our understanding of the epoch of reionization. Different mechanisms have been proposed to explain this phenomenon, including hot Wolf–Rayet stars (e.g. D. R. Garnett et al. 1991; C. Kehrig et al. 2011; M. Shirazi & J. Brinchmann 2012; A. Roy et al. 2025), shocks (e.g. A. Plat et al. 2019; L. M. Oskinova & D. Schaerer 2022), X-ray sources (e.g. P. Senchyna et al. 2020; C. Kehrig et al. 2021; C. Simmonds et al. 2021), and peculiar (nearly) metal-free hot stars,

PopIII(-like) stars (e.g. D. Schaerer 2002; P. Cassata et al. 2013; C. Kehrig et al. 2015, 2018; T. Nanayakkara et al. 2019; E. Pérez-Montero et al. 2020; A. Venditti et al. 2024).

Observations indicate an anticorrelation of the HeII line intensity with metallicity (e.g. N. G. Guseva et al. 2000; M. Shirazi & J. Brinchmann 2012), while HeII-emitters are expected to be more frequent in the high-z Universe (e.g. D. Schaerer 2002; E. R. Stanway & J. Eldridge 2019). This trend aligns with theoretical predictions indicating that lower metallicity galaxies exhibit harder ionizing spectral energy distributions (SEDs), more frequent in early-Universe conditions (e.g., R. Smit et al. 2015; E. R. Stanway & J. Eldridge 2019). Recent rest–frame–UV surveys have indeed uncovered dozens of strong HeIIλ1640 emitters at $2 \lesssim z \lesssim 6$ (e.g. P. Cassata et al. 2013; T. Nanayakkara et al. 2019; A. Saxena et al. 2020). Yet, despite deep observations, the large distances preclude spatially resolved studies and hinder multi-wavelength diagnostics, so the detailed physical conditions in these high-z systems remain poorly constrained.

Local analogues of high-z ($z \gtrsim 2$) star-forming galaxies (including those that dominated the epoch of reionization ($z \gtrsim 6$)) provide crucial laboratories for studying the properties of the high ionization gas thanks to their higher apparent magnitudes and spatial resolution. The SF dwarf galaxy IZw18 is a remarkable analogue, not only for high-z star-forming galaxies in general, but for a particular subset of these galaxies that present HeII emission, a feature frequently seen in LyC leakers (e.g., E. Vanzella et al. 2020; R. P. Naidu et al. 2022; D. Schaerer et al. 2022; A. U. Enders et al. 2023; C. Mondal et al. 2025). This is because of its extremely low metallicity [$12 + \log(O/H) = 7.11$, ∼3% solar; (e.g., J. M. Vílchez & J. Iglesias-Páramo 1998; C. Kehrig et al. 2016)] and also due to its highly extended nebular HeII-emitting region (∼ 5" diameter ∼ 400 pc at the distance of 18.2 Mpc; A. Aloisi et al. (2007)) which is spatially coincident with the NW SF knot, and with a corresponding very high HeII luminosity ($1.12 \pm 0.07 \times 10^{38}$ erg s$^{-1}$). These features were unveiled for the first time in C. Kehrig et al. (2015) based on integral field spectroscopic (IFS) data who find that the observed HeII-ionization budget of IZw18 can be explained by peculiar hot (nearly) metal-free massive stars (see also C. Kehrig et al. (2018)). In C. Kehrig et al. (2021), the authors discard X-ray photons as the dominant HeII ionizing mechanism in IZw18 highlighting the complex nature of HeII emission in this galaxy which keeps challenging state-of-the-art models (e.g., E. R. Stanway & J. Eldridge 2019; J. J. Eldridge & E. R. Stanway 2022). Additionally, R. J. R. Vaught et al. (2021) revealed notable velocity offsets (∼30 km s$^{-1}$) between the HeII-emitting gas and the bulk of the ionized gas traced by Balmer lines in IZw18 (see also A. Arroyo-Polonio et al. (2024)), suggesting dynamic decoupling and adding complexity to our understanding of the HeII ionization problem.

Regarding the presence of other high ionizing lines in IZw18, X-ray observations marginally revealed the extremely high-ionization line OVIII in IZw18 (D. Bomans & K. Weis 2002; T. X. Thuan et al. 2004), which indicates the presence of photons with energies above 59 Ryd. Nevertheless, T. X. Thuan & Y. I. Izotov (2005) searched for the high-ionization [NeV]λ3426 emission line (ionization potential of 97.1eV) in several SF dwarfs and did not detect such line in IZw18 (see also Y. Izotov et al. (2021)).

Additionally, H. Atek et al. (2009) identified pronounced Lyman α (Lyα) absorption in the northwest region of IZw18, close to the HeIIλ4686 emission. Their radiative transfer models pointed to a high neutral hydrogen column density ($N_{\rm HI} \simeq 6.5 \times 10^{21}$cm$^{-2}$) significantly suppressing the escape of Lyα and LyC photons into the intergalactic medium, even in conditions of minimal dust. Complementing this, several studies (e.g., R. Maiolino et al. 2017; S. R. Flury et al. 2023, 2024; A. Arroyo-Polonio et al. 2024; C. A. Carr et al. 2025) emphasize the critical role of energetic outflows in shaping nebular-line profiles, often manifesting as broad, blueshifted wings. Such outflows can carve low-density pathways facilitating LyC photon escape, highlighting that stellar-driven outflows play a vital role in photon leakage processes (e.g., R. Amorín et al. 2024).

In this letter, we take advantage of new optical [Gran Telescopio Canarias (GTC)/MEGARA; A. G. Gil de Paz et al. (2016)] and mid-infrared [James Webb Space Telescope (JWST)/MIRI] IFS data to shed new light on the puzzling source(s) of high-energy ionizing continuum (>54 eV) in IZw18.

## 2. OBSERVATIONS

This letter combines GTC/MEGARA optical IFS observations with archival mid-infrared IFS data from JWST/MIRI, complemented by archival imaging data from JWST/NIRCam and the Hubble Space Telescope (HST).

Optical observations of IZw18 were conducted on 2022 February 6 using the MEGARA instrument (LR-B grating; R ∼ 6000) at the Gran Telescopio Canarias (GTC), located at the Roque de los Muchachos Observatory (Spain). The Integral Field Unit (IFU) of MEGARA provides a field of view (FoV) of 12.5 × 11.3 arcsec$^2$, composed of 567 hexagonal spaxels (0.62 arcsec per spaxel), covering a spectral range of approximately over ∼ 4330–5200 Å. Three exposures of 1200 s were obtained on 2022-02-06, with a measured seeing of 1 arcsec during the observations; targeting the entire main body (MB) of IZw18 (R.A. = 09:34:01.98, Dec. = 55:14:26.5, J2000.0). The data reduction was performed following the procedure described in A. Arroyo-Polonio et al. (2024). The resulting data enable the generation of detailed maps of the structure and kinematics, particularly of the HeII$\lambda$4686 and H$\beta$ emission lines.

Moreover, we used archival mid-infrared IFU data from the MIRI Medium Resolution Spectrometer (MRS) of the JWST. These observations targeted the northwest knot of IZw18 and were conducted between March 8 and 9, 2024, as part of JWST Cycle 2 (PI: A.Aloisi, program 03533). We used all three MRS grating settings (SHORT, MEDIUM, LONG) within channel 3, covering wavelengths from ∼11.5 to 18 $\mu$m. This channel has a field of view of 5.2 ×5.5 arcsec$^2$, with a spectral resolving power ranging approximately from $R \sim 3,000$ at 12 $\mu$m to $R \sim 2,000$ at 18 $\mu$m. The spatial resolution is diffraction-limited, varying from ∼0.4 arcsec at the shortest wavelengths to ∼0.6 arcsec at the longest wavelengths. The full MIRIFULONG detector array operated in SLOWR1 readout mode with a 4-point dither pattern optimized for extended emission, achieving a total integration time of ∼29,146 s. The retrieved data are Level 3 science-ready calibrated products. The combination of these optical and MIR IFS observations carried out in this work allowed us to analyze the galaxy structure and ionization covering from one Rydberg to high ionization lines such as [NeV]14.32$\mu$m (ionization potential of ∼ 7.1 Ryb).

Additionally, we make use of archival imaging observations from JWST/NIRCam obtained with the F115W filter (PI: M. Margaret, $\lambda_c$ ∼1.15 $\mu$m), as well as HST data acquired with ACS/WFC using the F606W filter (PI: A. Aloisi, $\lambda_c$ ∼0.606 $\mu$m) and WFC3/UVIS with the F225W filter (PI: G. Oestlin, $\lambda_c$ ∼0.225 $\mu$m).

Some of the data presented in this article were obtained from the Mikulski Archive for Space Telescopes (MAST) at the Space Telescope Science Institute. The specific observations analyzed can be accessed via doi: 10.17909/yjfs-gc89.

## 3. DATA ANALYSIS

### 3.1. *Optical GTC/MEGARA IFS*

We investigated the spatial distribution, structure and kinematics of the ionized gas in the MB of IZw18 by analyzing two emission lines: H$\beta$ and HeII$\lambda$4686. These lines, each requiring different ionization potentials (13.6 eV for H$\beta$ and 54.4 eV for HeII$\lambda$4686), allowed us to sample the ionization structure within the ionized region. H$\beta$ maps the HII region and HeII$\lambda$4686 isolates an extended highly ionized zone (the HeIII region). By comparing these differently ionized layers, we gain a clearer, ionization-dependent view of the ionized gas in the galaxy.

For each spaxel within the GTC/MEGARA data cube, we fit single Gaussians to the emission lines to measure line flux, velocity, and velocity dispersion, following the relativistic Doppler formula and velocity dispersion correction described in A. Arroyo-Polonio et al. (2024), which subtracts the instrumental and thermal broadening ($\sigma_{\mathrm{inst}}$ and $\sigma_{\mathrm{thermal}}$) from the observed width so that $\sigma^2 = \sigma_{\mathrm{obs}}^2 - (\sigma_{\mathrm{inst}}^2 + \sigma_{\mathrm{thermal}}^2)$. A Gaussian smoothing kernel of FWHM = 1 arcsec (see top-left panel in Fig. 1) is applied to enhance the signal-to-noise ratio (S/N) and coherence between spaxels (A. Arroyo-Polonio et al. 2024). We excluded spaxels if their emission-line FWHM is below the instrumental FWHM ($\sigma_{\mathrm{inst}} = 0.45$ Å) or if the S/N falls below 3. Finally, all velocities are reported relative to the systemic velocity determined from the integrated H$\beta$ profile of the entire galaxy.

The maps of H$\beta$ and HeII$\lambda$4686 (see Fig. 1) reveal notable differences in the morphology and kinematics of the ionized gas in the MB of IZw18.

The HeII$\lambda$4686 flux map shows an extended emission across the northwest knot similar to the one seen in H$\beta$ (left panels in Fig. 1), but the peak of the two emission lines are displaced 1.6 arscec (140 pc projected at the distance of 18.2 Mpc; see also C. Kehrig et al. (2015)). This offset implies that the bulk of the ionizing sources producing photons with energies >1 Ryd (traced by H$\beta$) does not coincide spatially with the harder subsample of sources responsible for photons of >4 Ryd (traced by HeII$\lambda$4686). The gas structure of the HeIII region is not necessarily the same as that of the overall HII region. Consequently, the brightest sources in the HII region (as traced by HI recombination lines) does not necessarily trace the hard radiation fields. Furthermore, this spatial offset might suggest a scenario of star-formation propagation; however, given the still unclear nature of the HeII ionizing sources, such interpretation remains somewhat speculative. In addition, the velocity structure (middle panels in Fig. 1) in the HeII$\lambda$4686 region does not match the overall rotational pattern as H$\beta$ does. Furthermore, at the position of the H$\beta$ peak, the HeII$\lambda$4686 emission line is redshifted by ∼ +20 km s$^{-1}$ compared to the velocity of the H$\beta$ gas (similar results are found in R. J. R. Vaught et al. (2021)). This redshift might indicate that the HeII$\lambda$4686 emitting gas is situated deeper along the line of sight and/or is sig-

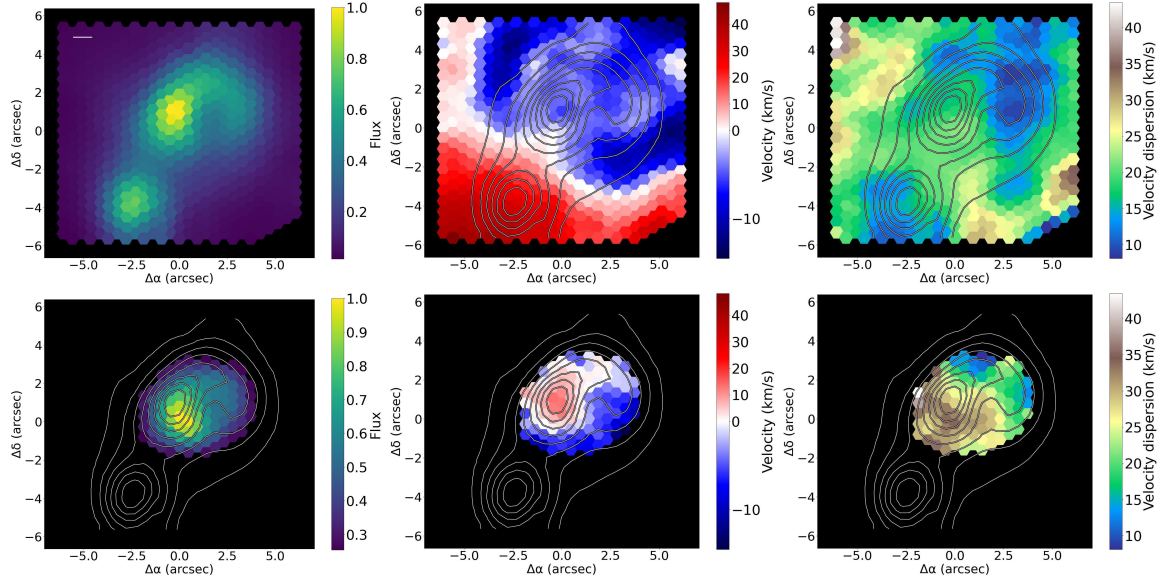

**Figure 1.** Maps of the flux (left), radial velocity (middle) and radial velocity dispersion (right) of the selected emission lines: Hβ (top), and HeIIλ4686 (bottom). The isocontours overplotted in all panels (except for the top-left one) represent the Hβ flux for reference. The white bar represented in the top-left panel represents 1 arcsec (i.e. ∼ 88 pc) which is the FWHM of the Gaussian kernel used to smooth the data. Black spaxels represent those that do not follow our criterion (S/N>3 and $\sigma > \sigma_{ins}$). In all figures in the letter North is up and East is to the left.

nificantly influenced by localized feedback mechanisms. Regarding the velocity dispersion maps (right panels in Fig. 1), the HeIIλ4686 line shows significantly higher velocity dispersions, being above 25 km s$^{-1}$ in a great portion of the region (meanwhile Hβ is below 25 km s$^{-1}$ in the HeII-emitting region) and exceeding 40 km s$^{-1}$ in certain areas. This broader linewidth suggests that the HeIIλ4686 emitting gas experiences additional kinetic energy input. The physical processes behind this could be shocks, intense stellar feedback or enhanced turbulence in highly energetic environments. (e.g., R. L. Davies et al. 2017; D. S. Rupke 2018; X. Yu et al. 2019)

We also extracted an integrated spectrum by adding the emission from all HeII-emitting spaxels with S/N > 3 in the HeII flux (see bottom left pannel in Fig. 1) with the aim to perform a detailed analysis of the line profiles shown in Fig. 2.

For the line profiles of both, HeIIλ4686 and Hβ, a Gaussian fit is made along with the velocity of the center of mass ($v_{CM}$), retrieved by using the equation:

$$v_{CM} = \frac{\sum_i F_i \times v_i}{\sum_i F_i}$$

where $F_i$ is the flux in each spectral pixel and $v_i$ is the velocity corresponding with the same spectral pixel. The $v_{CM}$ provides a mass-weighted average velocity of the emitting gas, offering a physically motivated estimate of the motion of each ionized region (HII and HeIII regions). It is particularly useful in cases where the line is asymmetric or shows broadening due to outflows, as it reflects the net kinematic behavior of the gas. The range of integration in order to retrieve the $v_{CM}$ is taken from -165 km s$^{-1}$ to 165 km s$^{-1}$ (4 times the FWHM of the HeIIλ4686 line profile). This choice ensures that the vast majority of the line emission, including possible extended wings due to turbulent or outflowing gas, is included in the computation while minimizing the contribution from noise-dominated continuum regions. To quantify the uncertainties associated with all these measurements, we employed the bootstrapping method (B. Efron & R. Tibshirani 1985) (see Table 1).

In agreement with the spatially resolved results, the line profiles clearly show that Hβ has a lower velocity dispersion in comparison to HeIIλ4686. Furthermore, the HeIIλ4686 peak is redshifted in comparison with the Hβ peak.

Beyond simply corroborating the earlier findings (higher velocity dispersion and redshifted peak in the HeIIλ4686 line relative to Hβ), the integrated spectrum reveals a new feature in the HeIIλ4686 line profile, a subtle blue-wing extension (bluer than -100 km s$^{-1}$) not clearly discernible in the spatial maps. This asymmetric line shape may indicate the presence of a blueshifted outflow (e.g., R. Amorín et al. 2024) in the highly ionized HeIIλ4686 region. To quantify this asymmetry we can compare the Gaussian centroid with $v_{CM}$. The higher the difference between these two the greater the asymmetry in the profile. As shown in Table 1, the Gaus-

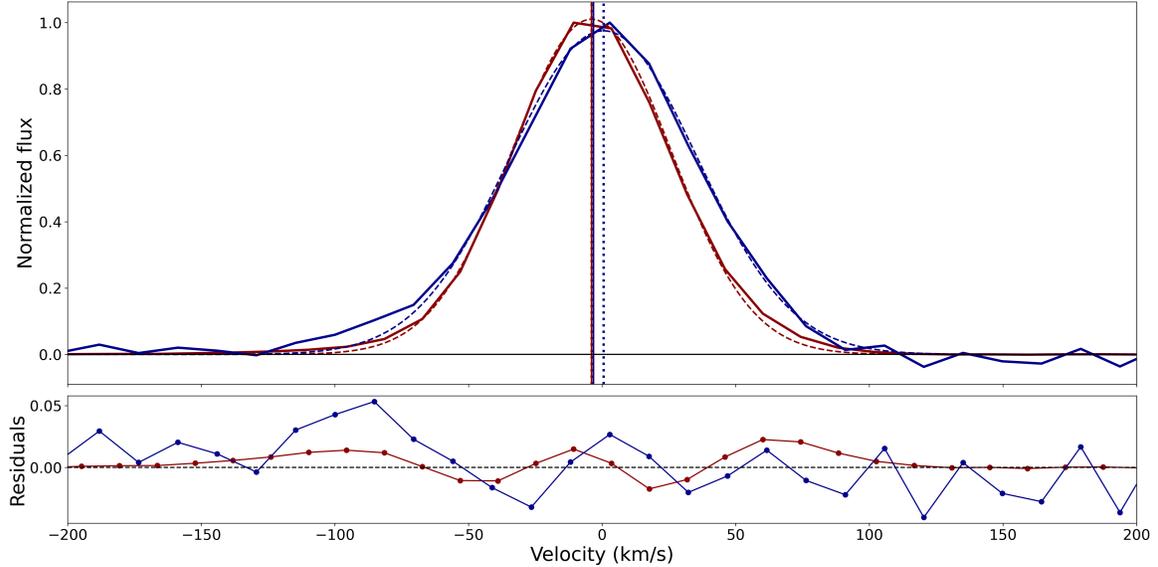

**Figure 2.** Line profiles of H$\beta$ (in red), and He II$\lambda$4686 (in blue). In the upper part, solid lines represent the observed line profiles and dashed lines the best Gaussian fit of each line. Vertical dotted lines represent the centroid of each Gaussian. Vertical solid lines represent the center of mass of each profile. In the bottom part the residuals (observation - Gaussian fit) are represented. The flux of each line is normalized to its peak value.

**Table 1.** Kinematic properties of the emission lines in the integrated spectrum.

| Emission line | Velocity (km s$^{-1}$) | Velocity dispersion ($\sigma$) (km s$^{-1}$) | $\mathbf{v_{CM}}$ (km s$^{-1}$) |
|---|---|---|---|
| (1) | (2) | (3) | (4) |
| H$\beta$ | -3.91±0.04 | 16.07±0.07 | -3.9±0.2 |
| He II$\lambda$4686 | 0.6±0.3 | 27.0±0.6 | -4±2 |

Column (1): Analyzed emission line. Column (2): Radial velocity. Column (3): Radial velocity dispersion. Column (4): Radial velocity of the center of mass.

sian centroid of H$\beta$ nearly coincides with its $v_{CM}$, both registering around 4 km s$^{-1}$. In contrast, while the He II$\lambda$4686 $v_{CM}$ remains close to 4 km s$^{-1}$, its Gaussian centroid is approximately 0 km s$^{-1}$.

The fact that H$\beta$ and He II$\lambda$4686 regions share the $v_{CM}$, together with the asymmetry in the He II$\lambda$4686 profile (characterized by a blue-wing and a redshifted peak) opens up the possibility of an off-centered feedback mechanism in the He III region, which is pushing a small fraction of gas with high velocity towards the blue. The conservation of momentum thus requires the other bigger fraction of He III gas to move towards the red, as seen in the redshift peak. This may indicate an early-stage outflow [some related studies investigating the inner outflow region and ionization-dependency of the outflow are M. S. Westmoquette et al. (2007); J. Chisholm et al. (2016)] that, depending on its kinetic energy and the mass it will encounter through its way out, could carve a path for ionizing photons that may eventually escape. We estimated the mechanical energy injected into the He III region by adopting a radius of $R = 200$ pc (i.e. a 400 pc diameter), a mean electron density of $n_e = 100$ cm$^{-3}$, a volume filling factor of $f = 0.1$, and an observed velocity dispersion of $\sigma = 30$ km s$^{-1}$ (Table 1). Under these conditions, the kinetic energy is

$$E_{\rm kin} = \frac{3}{2}\, M\, \sigma^2, \quad M = (\mu\, m_{\rm H}\, n_{\rm e}) \left(\frac{4}{3}\pi R^3\right) f,$$

where $\mu \simeq 1.4$ is the mean molecular weight of primordial H/He gas and $m_{\rm H}$ is the proton mass. Substituting the above values yields $E_{\rm kin} \simeq 3 \times 10^{53}$ erg, an order of magnitude result consistent with STARBURST99 for a $10^6\, M_\odot$ young star cluster (C. Leitherer et al. 1999). This mass compares well with the stellar mass of the NW knot (e.g., F. Annibali et al. 2013), suggesting that feedback from this knot could comfortably power the observed dispersion in velocity. Nevertheless, recent observations indicate that IZw18 exhibits significant opacity to both LyC and Ly$\alpha$ photons (H. Atek et al. 2009).

## 3.2. *mIR JWST/MIRI IFS*

Until now, [NeV] emission have never been observed in the galaxy IZw18. In this work, using IFS observations from JWST/MIRI, we report the detection of the [NeV]14.32$\mu$m line in IZw18 for the first time. This emission originates from an ion with ionization potential of 97.1eV, nearly twice as high as that of HeII$\lambda$4686 (see Fig. 3 and the right panel in Fig. 4).

Previous attempts to detect [NeV] emission in IZw18 (T. X. Thuan & Y. I. Izotov (2005) searching for the [NeV]3426Å line) were unsuccessful likely due to observational limitations. Our detection benefits from the superior spatial and spectral resolution of JWST/MIRI (approximately 0.5 arcsec and R$\sim$2400 at 14.3$\mu$m). Additionally, for the electron temperature and density of IZw18 [$T_e \sim 20,000$ K, $n_e < 300$ cm$^{-3}$; (e.g., C. Kehrig et al. 2016)], the emissivity of the [NeV]14.32$\mu$m line is roughly 70% higher than that of the [NeV]3426Å line [atomic data: CHIANTI v11 (R. Dufresne et al. 2024); computed with PyNeb (V. Luridiana et al. 2015)], significantly facilitating the detection of [NeV] in the IR.

To spatially correlate the JWST/MIRI and GTC/MEGARA cubes, we proceeded performing a careful spatial matching of the peak position of the H$\beta$ emission line from GTC/MEGARA with that of the Humphrey-$\alpha$ line observed with JWST/MIRI (see the right panel in Fig. 4). This precise spatial correlation of the H peaks allowed us to find that the peak emission of the [NeV]14.32$\mu$m line coincides very closely with the peak of the HeII$\lambda$4686 emission. Fig. 3 also shows that the [NeV] emitting region is extended [$\sim$ 1.8 arcsec (corresponding with a projected distance of $\sim$ 160 pc)] while covering a smaller area than that of the HeII-emitting region within the NW knot.

The nebular HeII$\lambda$4686 ionization (IP of 4 Ryd) in IZw18, still an open question, has been attributed to peculiar, nearly metal-free hot stars [popIII-like stars; see C. Kehrig et al. (2015, 2021)]. This new detection of the extended [NeV] emission (IP of 7.1 Ryd) and the spatial coincidence between the HeII$\lambda$4686 and [NeV] emitting zones have important implications for better understanding the high ionization sources in IZw18. These results not only reinforce the existence of a very hard SED within the NW knot (see how the [NeV] emission overlaps with some resolved stellar sources; right panel of Fig. 3) but also indicates the presence of much harder ionizing photons (E $>$ 97 eV is needed to produce NeV) (e.g., M. Mingozzi et al. 2025) which open the possibility of additional ionization mechanisms which will be studied in detail in a future work.

## 4. CONCLUSIONS

In this study, we conducted a detailed analysis of the ionized gas in the MB of IZw18 using the emission lines H$\beta$, HeII$\lambda$4686 and [NeV]14.32$\mu$m (each one tracing ionized gas at different ionization stages: 13.6 eV, 54.4 eV and 97.1 eV) using the IFS from GTC/MEGARA and JWST/MIRI. Our primary findings are summarized as follows:

1. The spatial distribution of H$\beta$ and HeII$\lambda$4686 emission lines reveals differences in their structure, with the HeII-emitting region peak offset by a projected distance of 140 pc from the peak H$\beta$ emission. This spatial separation indicates that the most extreme ionizing sources, responsible for the HeII emission, are concentrated away from the bulk of lower-energy (13.6 eV) ionizing sources traced by H$\beta$.

2. The kinematic analysis shows that HeII gas exhibits higher velocity dispersions and a different velocity pattern compared to that of the H$\beta$ emission. This indicates that the HeII emission region is kinematically decoupled from the HII region, possibly due to energetic processes such as localized stellar feedback or shocks, indicative of highly turbulent environments.

3. Integrated spectral analysis reveals an asymmetric, blueshifted extension in the HeII$\lambda$4686 profile, interpreted as evidence of early-stage stellar-driven outflow. Such outflow might potentially facilitate future ionizing photon leakage despite current observations of high Ly$\alpha$ opacity.

4. Using JWST/MIRI data, we report the first detection of the [NeV]14.32$\mu$m emission line in IZw18. The presence of [NeV] extended emission indicates ionizing sources with energies significantly exceeding those previously inferred solely from HeII emission. Additionally, the spatial coincidence between the peaks of [NeV] and HeII emissions suggests that the same highly energetic sources responsible for the [NeV] emission likely contribute to the bulk of HeII$\lambda$4686 emission.

Our results highlight the complexity and diversity of ionizing mechanisms at play within extremely metal-poor galaxies like IZw18, providing valuable insights into conditions analogous to those of early-Universe galaxies.

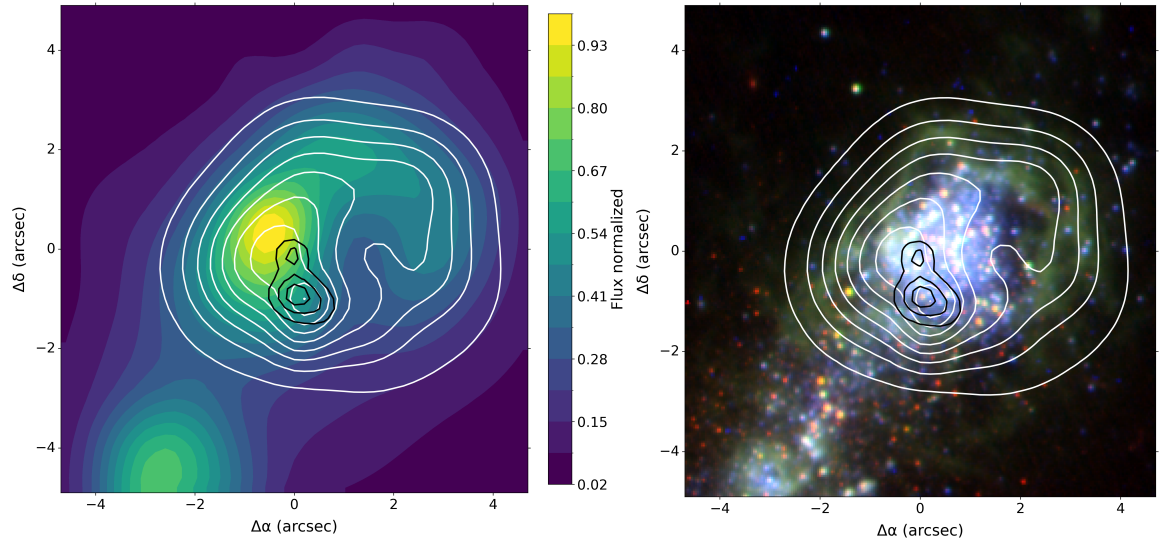

**Figure 3.** (Left panel) The MEGARA H$\beta$ emission map as a color-filled contour plot, smoothed using bilinear interpolation. For reference, the isocontours of the MEGARA HeII$\lambda$4686 and the MIRI [NeV]14.32$\mu$m emissions are overplotted in white and black, respectively. (Righ panel) Color-composite image in three bandpasses: near-infrared JWST NIRCam/F115W continuum at $\sim 1.15\mu$m (red), optical HST ACS/F606W continuum at $\sim 6060$Å (green), and HST WFC3/UVIS/F225W at $\sim 2250$Å (blue); the same isocontours of HeII$\lambda$4686 and [NeV]14.32$\mu$m line emissions are displayed.

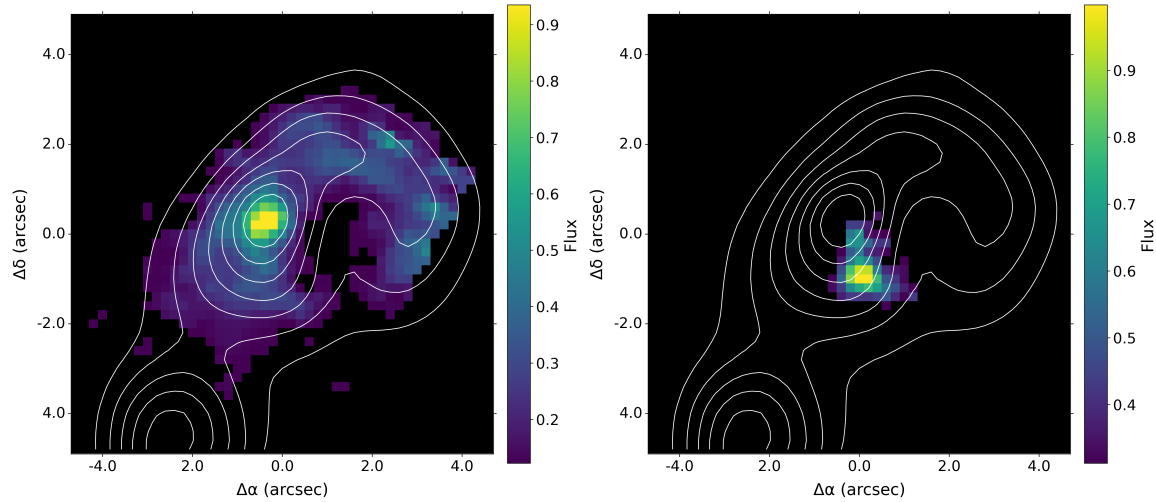

**Figure 4.** (Left panel) The MIRI Hu$\alpha$ emission map. For reference, the isocontours of the MEGARA H$\beta$ emission is overplotted in white. (Right panel) The MIRI [Nev]14.32$\mu$m emission map. Again, for reference, the isocontours of the MEGARA H$\beta$ emission is overplotted in white.

## ACKNOWLEDGMENTS

Author AAP acknowledges financial support from the State Agency for Research of the Spanish MCIU through Center of Excellence Severo Ochoa' award to the Instituto de Astrofísica de Andalucía CEX2021- 001131-S funded by MCIN/AEI/10.13039/501100011033, and from the grant PID2022- 136598NB-C32 "Estallidos8". Acknowledge support by the project ref. AST22_00001_Subp_11 funded from the EU – NextGenerationEU. The authors acknowledge the plan PID2021-123417OB-I00.